\documentclass[aps,prl,twocolumn,floatfix,showpacs,superscriptaddress]{revtex4-1}

\usepackage{amsmath,amsfonts}
\usepackage{graphicx}
\usepackage{color}
\usepackage{setspace}

\begin{document}

\title{Manifestation of axion electrodynamics through magnetic ordering on edges of topological insulator}

\author{Yea-Lee Lee}
\affiliation{Department of Physics and Astronomy, Center for Theoretical Physics, 
Seoul National University, Seoul 151-747, Korea.}

\author{Hee Chul Park}
\affiliation{Korea Institute for Advanced Study, Seoul 130-722, Korea.}

\author{Jisoon Ihm}
\email{jihm@snu.ac.kr}
\affiliation{Department of Physics and Astronomy, Center for Theoretical Physics, 
Seoul National University, Seoul 151-747, Korea.}

\author{Young-Woo Son}
\email{hand@kias.re.kr}
\affiliation{Korea Institute for Advanced Study, Seoul 130-722, Korea.}

\begin{abstract}
Based on a first-principles approach, we show that
in a single crystal of a prototypical topological insulator
such as Bi$_2$Se$_3$ the difference in the work function between adjacent surfaces 
with different crystal-face orientations generates a built-in electric field around facet edges.
Owing to the topological magnetoelectric coupling for a given broken time-reversal symmetry in the crystal,
the electric field, in turn, forces effective magnetic dipoles to accumulate along the edges,
realizing the facet-edge magnetic ordering.
We demonstrate that
the predicted magnetic ordering which depends only
on the work function difference between facets,
is in fact a manifestation of the axion electrodynamics in real solids.
\end{abstract}


\maketitle

A topological insulator (TI) hosts the topologically protected metallic surface states
on its boundaries between inner insulating bulk and outer vacuum,
that can exist on all the surfaces with different crystal orientations enclosing 
the crystal~\cite{review1,review2}.
Typically, the protected surface state
has the relativistic massless
dispersion relation around the time-reversal invariant momenta in the surface Brillouin zone
although its detailed features depend on surface characteristics~\cite{silvestrov,sidesurfacetheory1,
sidesurfaceexp,sidesurfacetheory2,jhlee}.
For example, the well-known TIs with the rhombohedral crystal structure
such as Bi$_2$Se$_3$, Bi$_2$Te$_3$ and Sb$_2$Te$_3$~\cite{biseries1,biseries2,ti5} have
stacked quintuple layers along the $(111)$ direction
and the low energy surface state on the $(111)$ surface
is isotropic in momentum space~\cite{biseries2} while
other surfaces
have quite anisotropic dispersions~\cite{silvestrov,sidesurfacetheory1,sidesurfaceexp,sidesurfacetheory2,jhlee}.
Besides changes in its low energy electronic dispersions,
different facets in a single crystalline TI would have many different physical
properties depending on their orientations
and the facet-dependent work function~\cite{AM,balderischi} is one of interesting examples among them.
In the TIs mentioned above,
such effects will be amplified because of its layered structure $-$
surface atomic and electronic
densities vary a lot depending on whether the surface is terminated along the layer or not.

Although the physical properties of 
topological states on a specific facet 
of three-dimensional (3D) TIs have been studied 
intensively~\cite{review1,review2,silvestrov,sidesurfacetheory1,sidesurfaceexp,sidesurfacetheory2,jhlee,
biseries1,biseries2,ti5},
mutual interactions among those 
contiguous to each other through edges
have not yet been examined well.
A trivial example is 
the coupling between two massless surface states on the opposite surfaces
resulting in an energy gap in the surface energy band of the TI thin film~\cite{tithinfilm}.
Even in a sufficiently large single 3D
TI crystal where the interaction between 
opposite surfaces can be neglected, 
different massless surface states should meet and interact with each other at
edges between two adjacent facets,
of which consequences have not been known well.
In this Letter, we show that the combining effects both from the usual surface-dependent
properties such as facet-dependent work function difference
and from the topological surface properties for a given broken time-reversal symmetry
produce a topological magneto-electric coupling (TME)~\cite{qah,sczhang2,sczhang1}
described by the axion electrodynamics
without external charge controls as conceived before~\cite{sczhang1}.
The resulting magnetic ordering along the edges should be robust and strong
enough to be measured.

\begin{figure}[b]
  \centering
  \includegraphics[width=0.95\hsize]{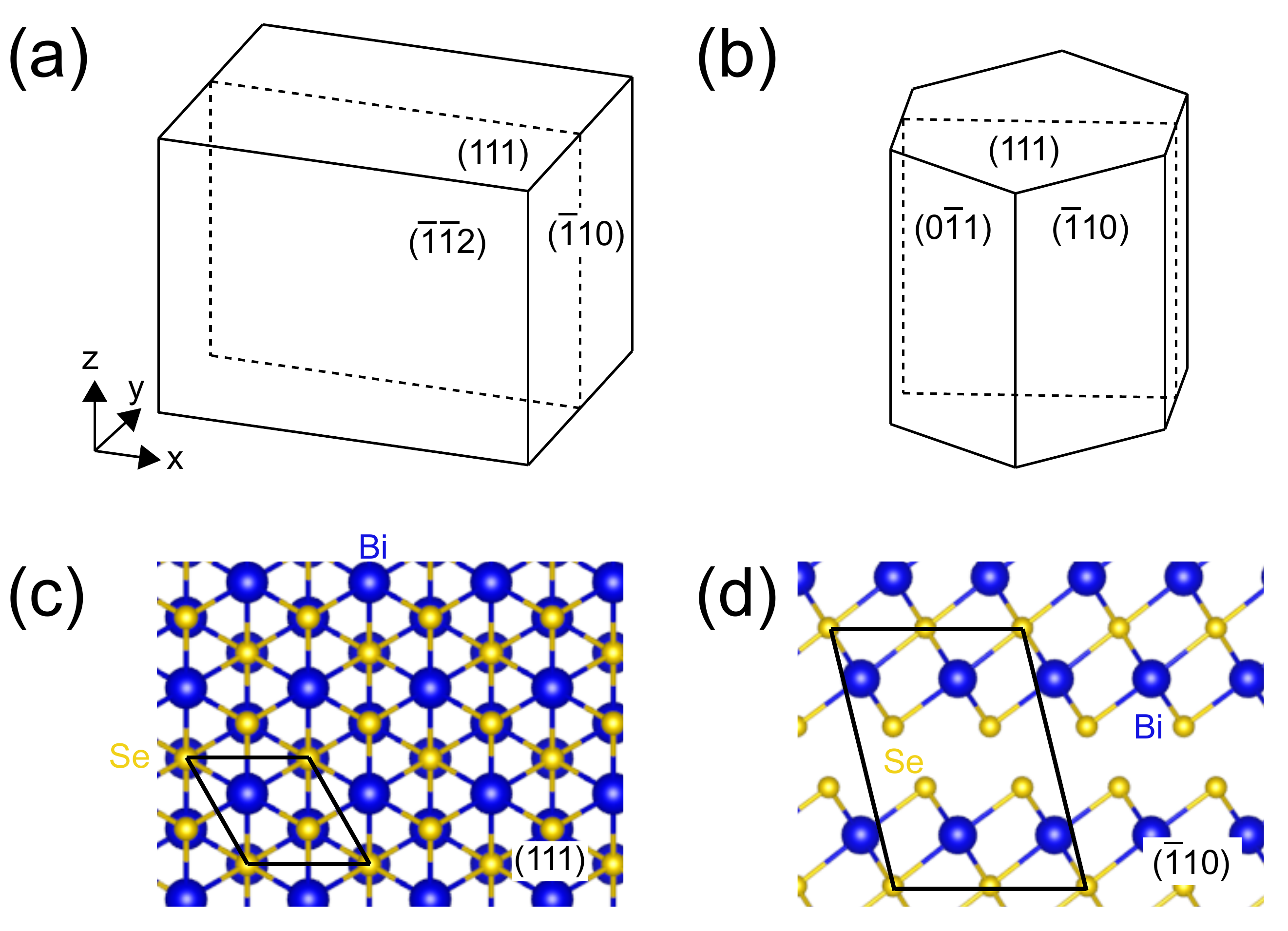}
  \caption{
(a) A rectangular shaped crystal of Bi$_2$Se$_3$ with the top ($111$), 
the front [$\bar{1}\bar{1}2$], and the side [$\bar{1}10$] surfaces. 
(b) A hexagonal one with the top ($111$) surface and the [$\bar{1}10$] sides.
The dashed squares in (a) and (b) are cross sections of the crystal 
and will be considered in the model calculations.
Atomic structures of (c) the ($111$) and (d) the ($\bar{1}10$) surfaces. 
The black parallelogram indicates the unitcell of each surface,
of which area for the $(111)$ surface is 14.83~\AA$^2$ while
for $(\bar{1}10)$ surfaces 68.42~\AA$^2$.
}
\end{figure}

We first examine the electronic structures of
topological surface states on various facets based on first-principles calculation methods~\cite{vasp}
with atomic pseudopotentials~\cite{paw} and the local density approximation~\cite{ca} 
including spin-orbit interactions.
We choose Bi$_2$Se$_3$ as an example material for our consideration [Fig. 1].
For Bi$_2$Se$_3$,
a surface with the $(111)$ direction has
a triangular lattice of Se atoms (typical cleavage surface)
while one with the $(\bar{1}10)$ or the $(\bar{1}\bar{1}2)$
direction perpendicular to the $(111)$ direction
has a tetragonal surface unitcell (Fig. 1)~\cite{lattice}.
The slab thickness is chosen as 6 quintuple layers 
and 15 monolayers for each (111) and ($\bar{1}$10) surface
where the interaction between top and bottom surfaces is ignorable.
To reduce spurious interaction
between the neighboring slabs we introduce a large vacuum over 20~\AA.
In a single crystal of Bi$_2$Se$_3$ grown along the $(111)$ direction,
the rectangular shaped crystal has the $[\bar{1}10]$ and $[\bar{1}\bar{1}2]$
surfaces as side walls
while hexagonal~\cite{chem} or triangular~\cite{epit} column shaped one
has the $[\bar{1}10]$ surfaces as side surfaces as shown in Figs. 1(a) and (b).

The electronic band structures indicate
that the protected massless metallic surface state
on the $(\bar{1}10)$ surface has a quite anisotropic
dispersion relationship
unlike the well known isotropic one on the $(111)$ surface [Fig.2].
In addition to changes in the shape of the dispersion,
the energy level of the Dirac point of the surface state
with respect to the bulk chemical potential shifts
depending on the surface orientation owing to intrinsic anisotropy of the system.
Our estimation of the shift within slab geometry calculations
is 120 meV for Bi$_2$Se$_3$ that is comparable to a previous estimation~\cite{silvestrov}.

\begin{figure}
  \centering
  \includegraphics[width=1.0\hsize]{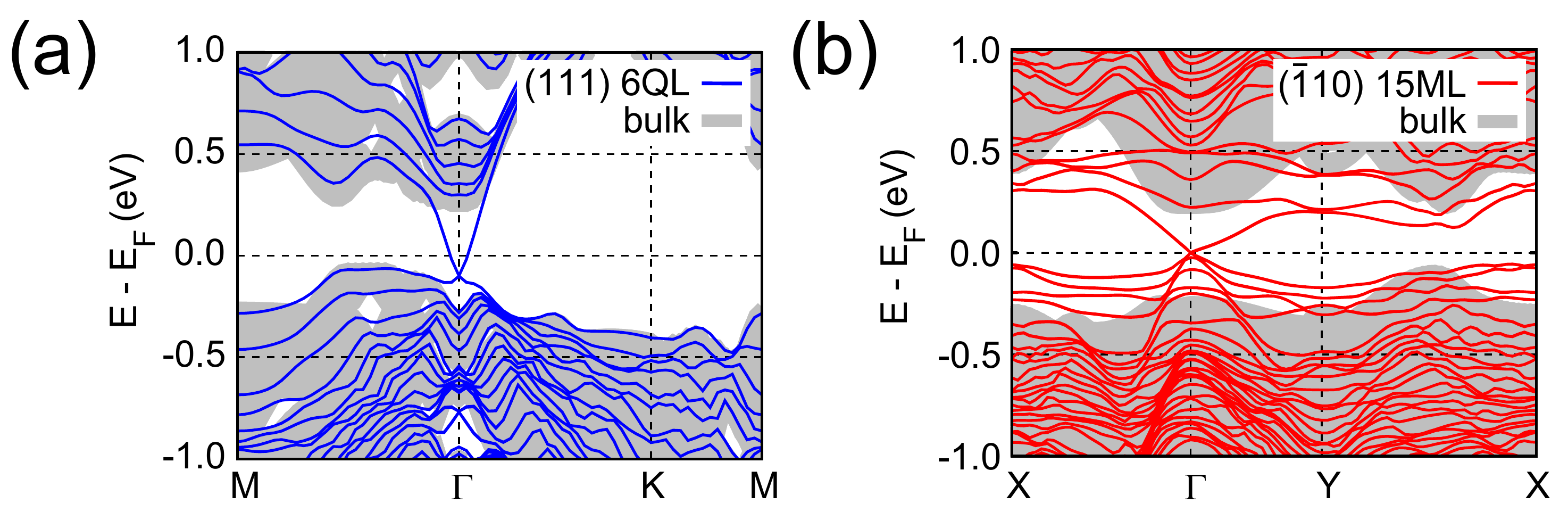}
  \caption{
The band structures of the ($111$) surface of 6 quintuple layer (QL) (a)
and the ($\bar{1}10$) surface of 15 monolayer (ML) (b) on energy-momentum space.
Projected bulk states are shown by a gray region for both surfaces.
}
\end{figure}

Next, we study the work function difference between different facets
of Bi$_2$Se$_3$.
The work function is the energy required to extract
an electron from a specific surface of a solid and is determined by the difference
between the vacuum and the Fermi level~\cite{AM} and 
obtained by the energy difference
between the potential energy on the vacuum and the Fermi level in our slab calculations.
The calculated work function of the $(111)$ surface
is $W_{\text{(}111\text{)}}=5.84$ eV and that of the $(\bar{1}10)$ surface 
$W_{\text{(}\bar{1}10\text{)}}=5.04$ eV.
Full relaxation of atomic structures does not change the calculated
values. We also find that the work function
of the $(\bar{1}\bar{1}2)$ surface is 5.11 eV, similar
to that of the $(\bar{1}10)$ surface.
Such a large work function difference mainly
originates from the surface-dependent dipole moments
when the inversion symmetry is broken at the surface,
and it has also been well-known in previous first-principles calculations on elemental
metals such as Aluminum~\cite{balderischi}.

\begin{figure}
\includegraphics[width=1.0\hsize]{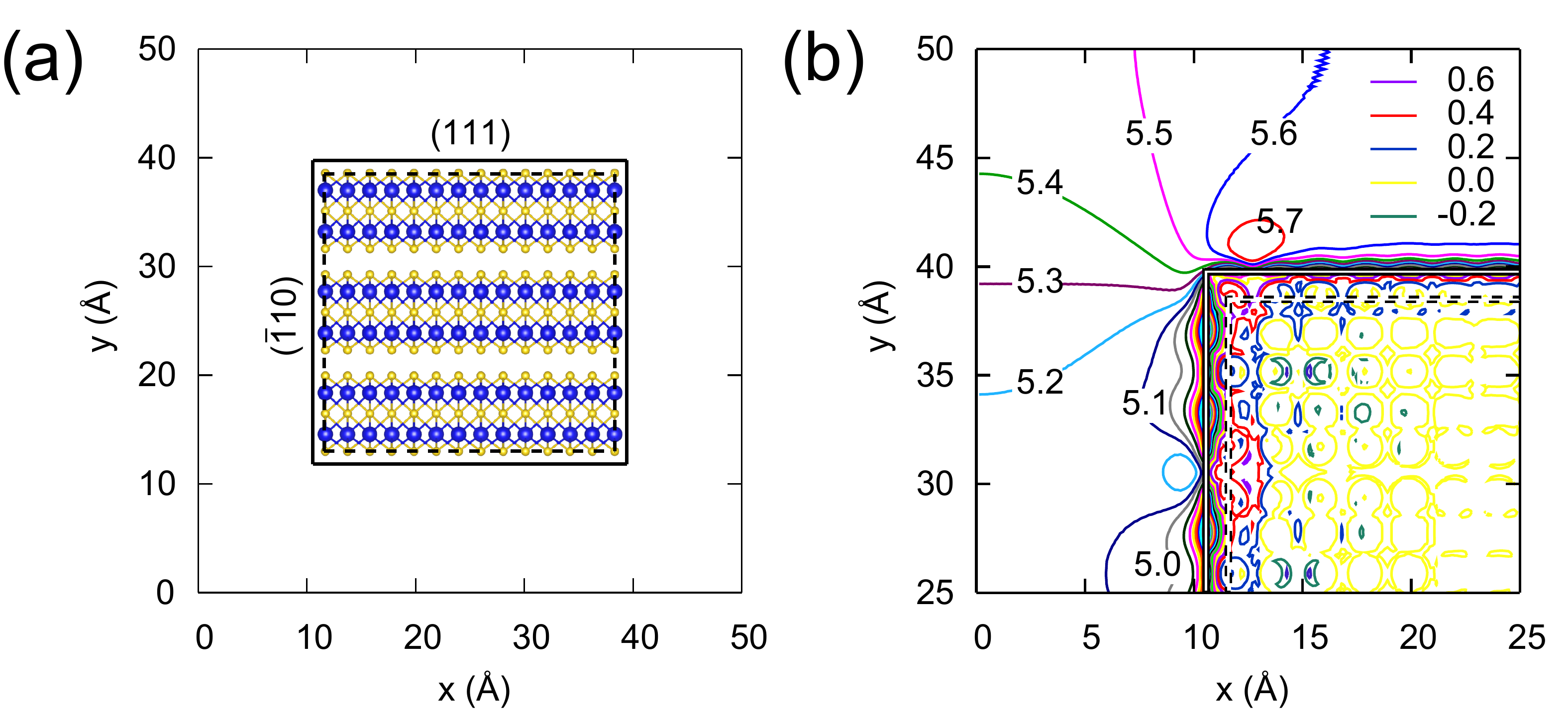}
\caption{\label{nanorod}
(a) The supercell structure of the nanorod with  $(110)$ and $(\bar{1}10)$ side surfaces. 
(b) Contour plot of the potential energy difference between the nanorod and the bulk
shown for left-upper side.
The contour interval is chosen as 0.2 eV for the contour values less than 5.0 eV, and 0.1 eV for those greater than 5.0 eV.
Dashed and solid lines in (a) and (b) denote the outermost atomic positions and the region of subtracting the bulk potential from the nanorod one, respectively.}
\end{figure}

While the electrostatic potential variation generated by work function differences
occurs only outside the bulk and near edges due to metallic screening in elemental metals~\cite{balderischi},
the TI supports the variation inside as well as outside the bulk
with appropriate dielectric screenings.
The Hartree potential distribution for the cross section of Bi$_2$Se$_3$ nanorod shown in Fig. 3(a)
can be obtained from actual \textit{ab-initio}
calculations. 
The resulting potential distributions exhibits
spatially varying electric potential in the vacuum region owing to the large 
work function difference between $(110)$ and $(\bar{1}10)$ sides of the nanorod
and a constant electric potential sufficiently ($>$ a few \AA) inside the bulk [Fig. 3(b)].
Near the surface (both inside and outside the bulk within 5 \AA), 
the rapid potential variation is also confirmed by this calculation.
Two boundary conditions, the calculated work function on the surfaces and 
a constant potential inside the bulk,
will be considered in the next model below.
%
%

Having established the electric potential distribution around the edges
of the 3D Bi$_2$Se$_3$, we now turn to its consequences on TME couplings~\cite{sczhang2}.
As long as the TI has an energy gap opening at the surface by breaking the time-reversal symmetry~\cite{sczhang2,sczhang1}
(achievable by depositing thin magnetic films~\cite{qah,mimp}, for example) and its Fermi level resides inside the gap,
the low energy physics can be described phenomenologically in terms of the axion electrodynamics~\cite{axion}.
In addition to the ordinary Maxwell Lagrangian~\cite{jackson}, the axion electrodynamics has an effective term
$L_{\text{eff}}= \kappa\theta\mathbf{E}\cdot\mathbf{B}$~\cite{axion}
where $\kappa=e^2/2\pi h$, $h$ is the Planck's constant, $e$ is the electric charge,
$\mathbf{E}$ is the electric field, $\mathbf{B}$ is the magnetic induction,
and $\theta$ is the axion field.
While the constant $\theta$ plays no role
in the equations of motion,
the gradient of $\theta$
gives rise to an extra charge density and a current density~\cite{axion,Essin}
that are related to the TME couplings.
Consequently, the electric displacement $\mathbf{D}$
and the magnetic field $\mathbf{H}$ can be
rewritten assuming a linear dielectric for Bi$_2$Se$_3$,
$\mathbf{D}=\epsilon \mathbf{E} \mp \kappa\theta\mathbf{B}$ and 
$\mathbf{H}=\frac{1}{\mu} \mathbf{B} \pm \kappa\theta\mathbf{E}$,
where the double signs for the axion field $\theta$ 
are in the same order~\cite{sczhang2,sczhang1}.
We introduce a magnetic ($\psi$) and electric ($\phi$) scalar potentials
such that the magnetic field 
$\mathbf{H}$ is obtained from $\mathbf{H}=-\nabla\psi$
and 
the electric field 
$\mathbf{E}=-\nabla\phi$.
With these scalar potentials, we can convert the electrodynamics into a variational
problem for both potentials $\delta F(\phi,\psi)\equiv\delta F_\textrm{M}(\phi,\psi)
+\delta F_\textrm{A}(\phi,\psi)=0$.
Here, $F_\textrm{M}=\frac{1}{2}\iint_{\Omega}
\left[\epsilon(\nabla\phi)^2+\mu(\nabla\psi)^2 \right]d\Omega$
and $F_\textrm{A}=\frac{1}{2}\iint_{\Omega}
\left[ \mu\kappa^2\theta^2(\nabla\phi)^2
\mp 2\mu\kappa\theta(\nabla\phi\cdot\nabla\psi) \right] d\Omega$,
where $\Omega$ is the domain area. 
$F_\textrm{M}$ corresponds to the ordinary Maxwell electrodynamics while 
$F_\textrm{A}$ to the axion term unique to the TI.
Then we solve it numerically using the finite element method~\cite{FEM} 
with a given boundary condition
for this model system~\cite{boundary}.

\begin{figure}
  \centering
  \includegraphics[width=0.6\hsize]{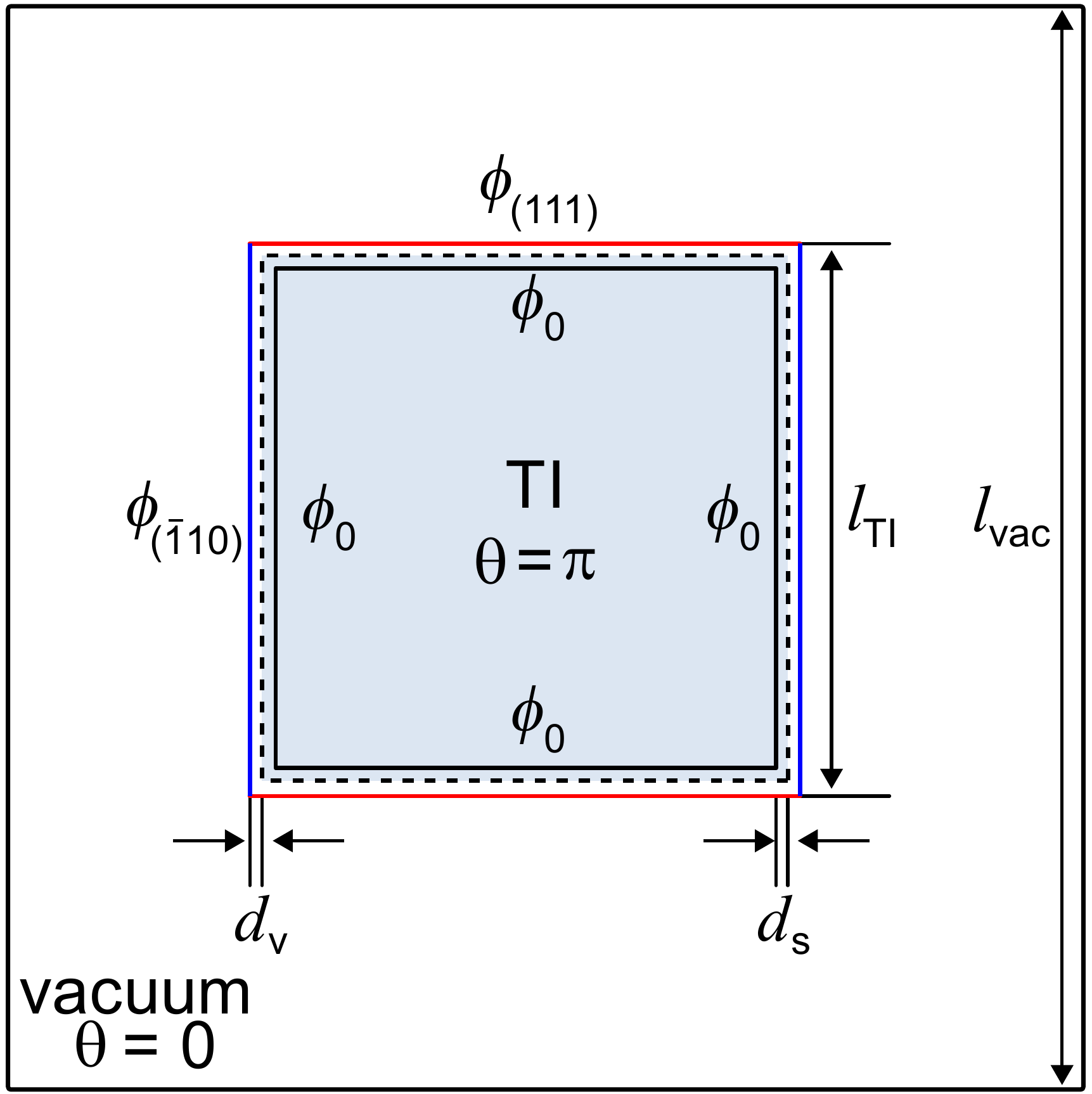}
  \caption{\label{model}
The cross section of a Bi$_2$Se$_3$ crystal as the model system.
In a vacuum box of $l_{\text{vac}}$ = 1 $\mu$m,  
the TI crystal (dashed-line box) of $l_{\text{TI}}$ = 100 nm locates at the center.
The two additional boxes for the boundary values lie on the TI crystal with the spacings, 
$d_{\text{s}}= d_{\text{v}}$ = 5 \AA, which are chosen from \textit{ab-initio} calculations.
The boundary condition is given by the fixed potentials of $\phi_{\text{(}111\text{)}}= -5.84$ V, 
$\phi_{\text{(}\bar{1}10\text{)}}= -5.04$ V and $\phi_0$ = 0 V on the red, blue, and black lines, respectively.
}
\end{figure}

We choose a rectangular cross section of the 3D TI crystal such as the dashed square in Fig. 1
to describe essential features of the low energy electrodynamics.
In Fig. 4, each side of the dashed-line box corresponds to 
the (111) and the ($\bar{1}10$) surfaces of Bi$_2$Se$_3$, respectively.
We set two additional boxes inside and outside the TI crystal with spacing of $d_{\text{s}}$ and $d_{\text{v}}$ 
for implying boundary conditions.
The electric potential on each side (outer box) is determined by the negative work function value of each facet divided by $e$,
such as
$\phi_{\text{(}111\text{)}}=-W_{\text{(}111\text{)}}/e = -5.84$ V (red line) 
and $\phi_{\text{(}\bar{1}10\text{)}}=-W_{\text{(}\bar{1}10\text{)}}/e = -5.04$ V (blue line).
Sufficiently inside the TI crystal the dielectric screening leads to the constant potential, so that
the potential of the inner box is fixed by zero ($\phi_0$ = 0 V).
Across the boundary between the dashed-line box to outer vacuum, 
the axion field and permittivity
change from $\pi$ and $100\epsilon_0$~\cite{dielectric} to 0 and $\epsilon_0$, respectively, while
the permeability ($\mu_0$) remains to be the same [Fig. 4].

A large $\mathbf{E}$ is given around the edges of Bi$_2$Se$_3$ at the beginning,
the numerical solution of $\delta F=0$ with the boundary conditions
gives $\mathbf{H}$ as well as $\mathbf{B}$ (and then $\mathbf{D}$) 
thanks to the axionic field $\theta$.
If the chemical potential of the TI crystal locates in the surface energy gap 
which are opening with a certain broken time-reversal symmetry operation
(e.g., 
the magnetization of the magnetic thin film~\cite{sczhang2,sczhang1}), 
we expect a strong TME effect around all the edges connecting two facets
with different orientations (hereafter we choose the upper sign for the $\theta$).


\begin{figure}
  \centering
  \includegraphics[width=1.0\hsize]{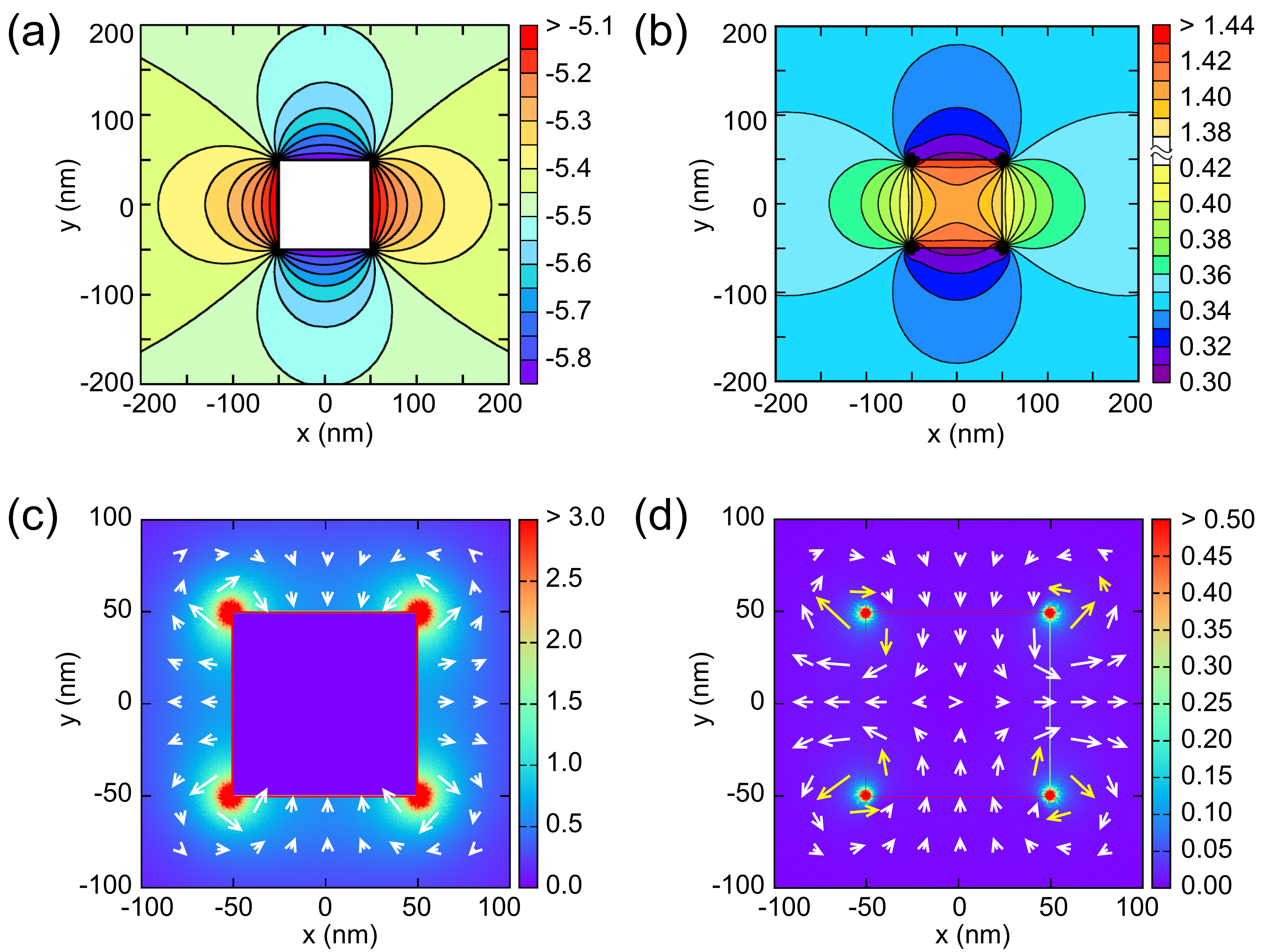}
  \caption{\label{main}
Contour plots of (a) the electric (V) 
and (b) the magnetic scalar potential ($10^{-6}$ C/s) for the model in Fig. 4.
The contour intervals are 0.05 V and $10^{-8}$ C/s for (a) and (b), respectively.
%
The color scale is jumped from -5.05 V (red) to zero (white) in (a) and from 0.42 $\times 10^{-6}$ C/s to 1.38 $\times 10^{-6}$ C/s
in (b) because of the huge potential change.
(c) The electric field (10$^{\text{7}}$ V/\AA) 
and (d) the magnetic induction (gauss) with strength denoted by colors and direction by arrows. 
%
The length of arrows is proportional to the strength of fields while
the yellow arrows are shortened by a half of the original size.
 }
\end{figure}

The contour plot of the electric and the magnetic scalar potentials
are achieved through the aforementioned procedures as shown in Figs. 5(a) and (b), respectively.
The electric potential rises continuously starting from the low-potential
(111) surface, while it goes through a maximum
to the high-potential ($\bar{1}10$) surface.
There is a discrete jump in the electric potential at edges where the two surfaces meet on the outer box boundary,  
and the field would diverge for an infinitely sharp edge.
However, we have confirmed that the locally-averaged field is practically independent of the sharpness of the edge, 
and smoothing the potential variation near the edge does not change our results~\cite{suppl}.
The magnetic scalar potential profile appears in the similar structure to the electric potential,
because the effective Lagrangian from the axion electrodynamics has
a symmetric structure between the electric and the magnetic parts.
Near edges the small change occurs in the magnetic scalar potential due to the boundary condition on the inner box
applied for the electron screening.
This change is not a critical fact to determine the overall physics in this system~\cite{suppl}.
For both electric and magnetic potentials, there are huge changes in the very narrow region near the surfaces, 
and the contour lines are very dense as shown in Figs. 5(a) and (b).
These huge potential gradients near the surface finally lead to the large field near the surfaces.

The electric field $\mathbf{E}$ and the magnetic induction $\mathbf{B}$
are given in Figs. 5(c) and (d), respectively.
Both fields are very high at edges and decrease
toward the center of each surface.
At 5 nm away from the edge, we obtain the electric field of $3.88\times10^7$ V/m
and the magnetic induction of 0.14 gauss.
These values are orders of magnitude higher than the previously predicted value
for the topological magnetic inductions induced by a point charge on a flat geometry~\cite{sczhang1}
because of enhanced geometric effects for edges as well as large intrinsic electrostatic potential
differences between two facets.
Outside the TI crystal, the electric field heads from the (111) surface for the ($\bar{1}10$) surface,
and the magnetic induction is parallel to the electric field.
The electric field inside the TI crystal near the surface is perpendicular to the each surface (outward direction) 
because of the boundary structure.
The magnetic induction follows the right-hand rule with respect to the quantized Hall current at edges 
which is given by the electric field and the gradient of the axion field such as
$\mathbf{j}_{\text{H}}=-\kappa(\nabla\theta\times\mathbf{E})$,
so that the asymmetric dipole-type magnetic induction arises at the edges~\cite{suppl}.



In conclusion, 
we have examined
electromagnetic properties of the single crystalline TI 
and showed that 
the large electric field and the associated topological magnetic ordering can be naturally generated 
by the facet-dependent work function near the edges of crystal without external manipulation.
Our demostration can be a useful basis
to realize the axion electrodynamics in solids and to study 
various other axionic aspects of TIs.
Furthermore, 
when the two necessary conditions, nonzero magnetoelectric coupling and gapped bulk spectrum, are fullfilled, 
the predicted edge magnetic ordering should occur
 because the facet-dependent work function difference is present in almost all crystals.
Thus, our mechanism is not limited to the TI crystals with broken time reversal symmetry
but could be extended to other materials without non-trivial Chern-Simon magnetoelectric coupling 
such as Cr$_2$O$_3$~\cite{souza}.




\begin{acknowledgements}
Y.-L.L. and H. C. P. contributed equally to this work.
We thank Y. Kim, Y. B. Kim, J. H. Han, B.-J. Yang, E.-G. Moon 
and S. Coh for fruitful discussions.
Y.-L.L. and J.I. were supported by the NRF funded by the MSIP of Korean government
(Grant No. 2006-0093853). 
Y.-W.S. was supported by the NRF funded by the MSIP 
of Korean government (CASE, 2011-0031640 and QMMRC, No. R11-2008-053-01002-0).
Computations were supported by the KISTI and the CAC of KIAS.
\end{acknowledgements}


%

\end{document}


\title{Supplemental Material for ''Manifestation of axion electrodynamics through magnetic ordering on edges of topological insulator''}

\author{Yea-Lee Lee}
\affiliation{Department of Physics and Astronomy, Center for Theoretical Physics, 
Seoul National University, Seoul 151-747, Korea.}

\author{Hee Chul Park}
\affiliation{Korea Institute for Advanced Study, Seoul 130-722, Korea.}

\author{Jisoon Ihm}
\affiliation{Department of Physics and Astronomy, Center for Theoretical Physics, 
Seoul National University, Seoul 151-747, Korea.}

\author{Young-Woo Son}
\affiliation{Korea Institute for Advanced Study, Seoul 130-722, Korea.}

\renewcommand*{\thefigure}{S\arabic{figure}}

\maketitle

\section{Detailed information for numerical results}
\subsection{The electric and magnetic properties very near the edge}

\begin{figure} [b]
\includegraphics[width=0.47\textwidth]{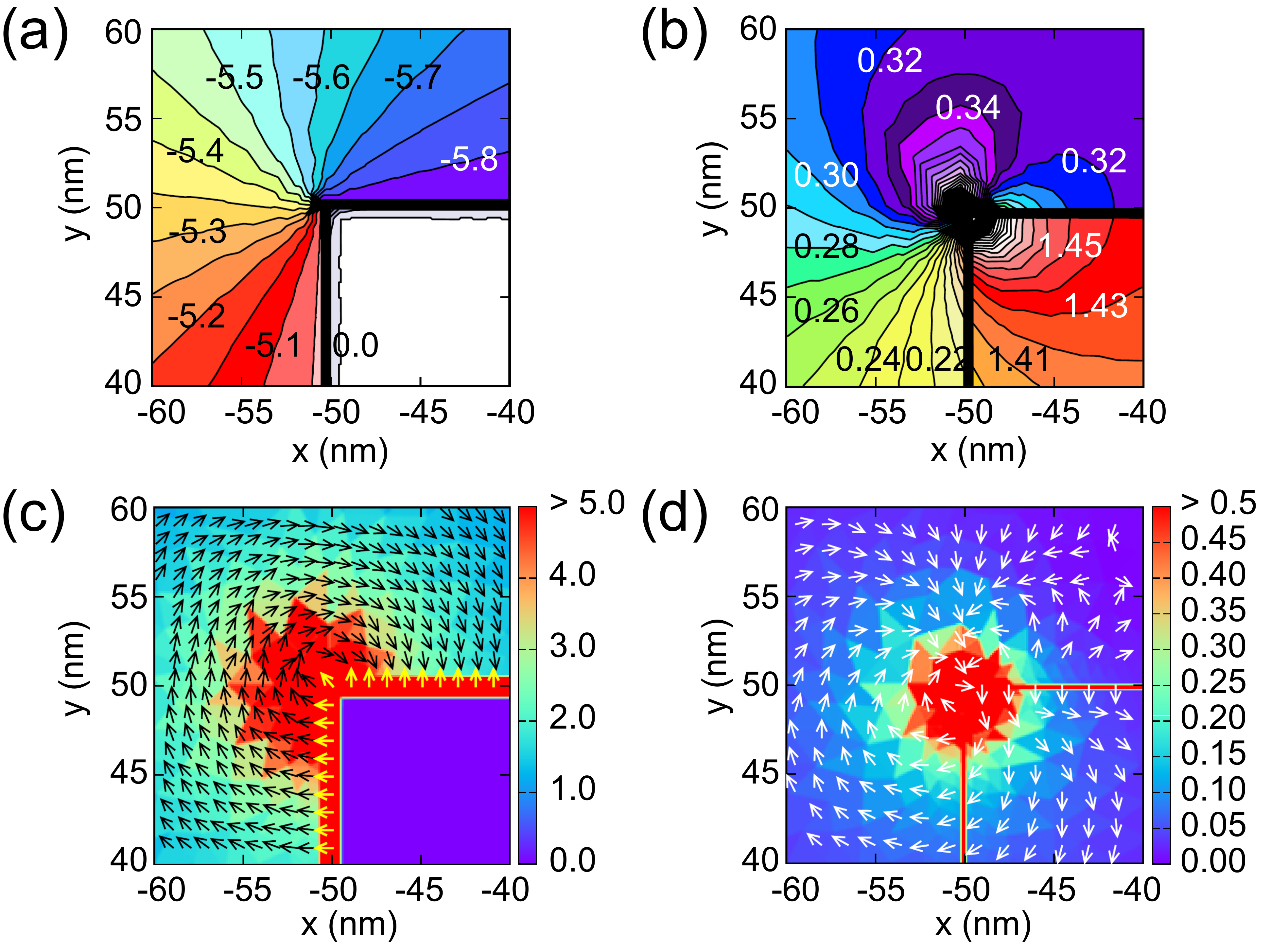}
\caption{\label{edge}
The electric and magnetic potentials and fields around the upper-left corner of the TI crystal.
The contour plots of the electric potential (V) and the magnetic scalar potential ($10^{-6}$ C/s) are shown in (a) and (b).
The contour lines are evenly spaced of 0.05 V and 10$^{-8}$C/s for both potentials, respectively, and coloring between the contour lines
helps the clearness of data.
The electric field (10$^{7}$ V/\AA) and the magnetic induction (gauss) are drawn in (c) and (d).
Colors denote the magnitude of the field at each point following the color bar on the right,
and directions of fields are represented by arrows whose lengths are normalized.
The electric field inside the TI crystal is emphasized by the yellow arrow in (c), 
and it is obtained by the dielectric screening.
Furthermore, the magnetic scalar potential and field are also affected by the screening effect, so that they are not symmetric to electric parts.
}
\end{figure}

In our model system, the direction of the electric field points from the ($\bar{1}$10) surface to the (111) surface outside the TI crystal, and 
the electric field directs from the inner to the outer box inside the TI crystal [Fig.~\ref{edge}].
The inner and outer boxes are shown in Fig. 4 in the main text, and the electric potential on each box is fixed by zero and negative work function of each surface divided by $e$, respectively.
There are two potential-fixed parallel plates near surfaces, and
the electric field follows the direction from higher electric potential (inner box)
to lower one (outer box).
From these reasons magnetic scalar potential and field are not symmetric to electric ones around the corner, 
and we can consider it as a screening effect in a dielectric media.
The butterfly-shaped magnetic field appears in Fig.~\ref{edge} (d), as if an asymmetric magnetic dipole lies on the corner.
It comes from the opposite direction of the quantized Hall current at the corner.
The quantized Hall current is $\mathbf{j}_{\text{H}}= -\kappa(\nabla\theta\times\mathbf{E})$,
and at the upper-left corner the electric field and the gradient of the axion field are given by 
$\mathbf{E}=-|\mathbf{E}_1|\hat{x}+|\mathbf{E}_2|\hat{y}$ (where $|\mathbf{E}_1|\sim |\mathbf{E}_2|$) and 
$\nabla\theta=\pi\hat{x}\delta(x-x_0) -\pi\hat{y}\delta(y-y_0)$, where ($x_0$,$y_0$) is the position of the corner.
The Hall current is obtained as 
$\mathbf{j}_{\text{H}} = (e^2/2h)\hat{z}(|\mathbf{E}_2|\delta(x-x_0)-|\mathbf{E}_1|\delta(y-y_0))$,
the opposite direction Hall current at the corner generates the dipole-like field structure.

\subsection{Electric and magnetic charge densities}

\begin{figure} [b]
\includegraphics[width=0.45\textwidth]{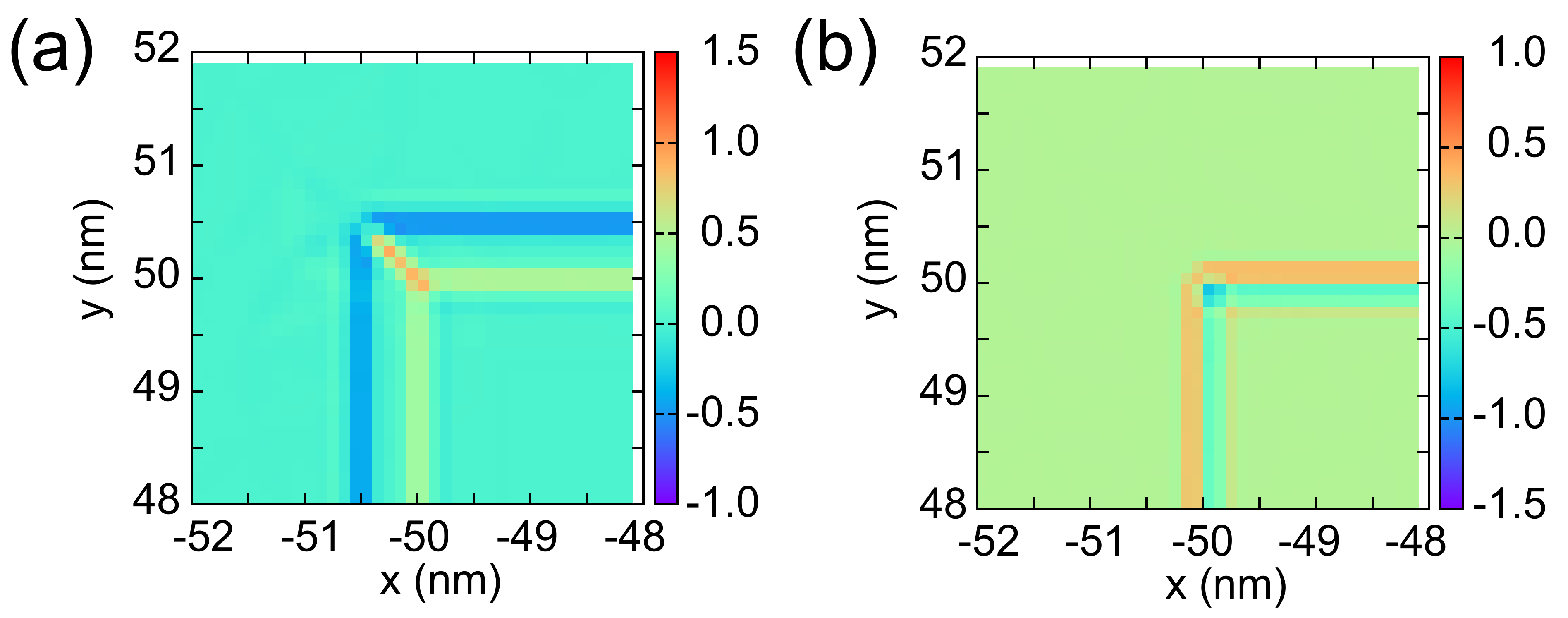}
\caption{\label{charge}
(a) The electric charge density ($10^{9}$ C/m$^2$) and (b) the magnetic charge density ($10^{15}$ C/ms) around the upper-left TI corner.}
\end{figure}

The derivatives of electric field and magnetic induction give rise to the electric and magnetic charge densities,
such as
$\nabla\cdot\mathbf{E}=\rho/\epsilon_0$ and $\nabla \cdot \mathbf{B} = \mu_0\rho_M$.
The electrons tend to move to higher electric potentials, so that electric density is positive (negative) on the inner (outer) box
shown in Fig.~\ref{charge} (a).
Magnetic charges appear on the surface where the axion field is changed (Fig. ~\ref{charge} (b)),
and the magnetic dipoles are ordered.

\subsection{ Smoothing boundary conditions}

\begin{figure}
\includegraphics[width=0.48\textwidth]{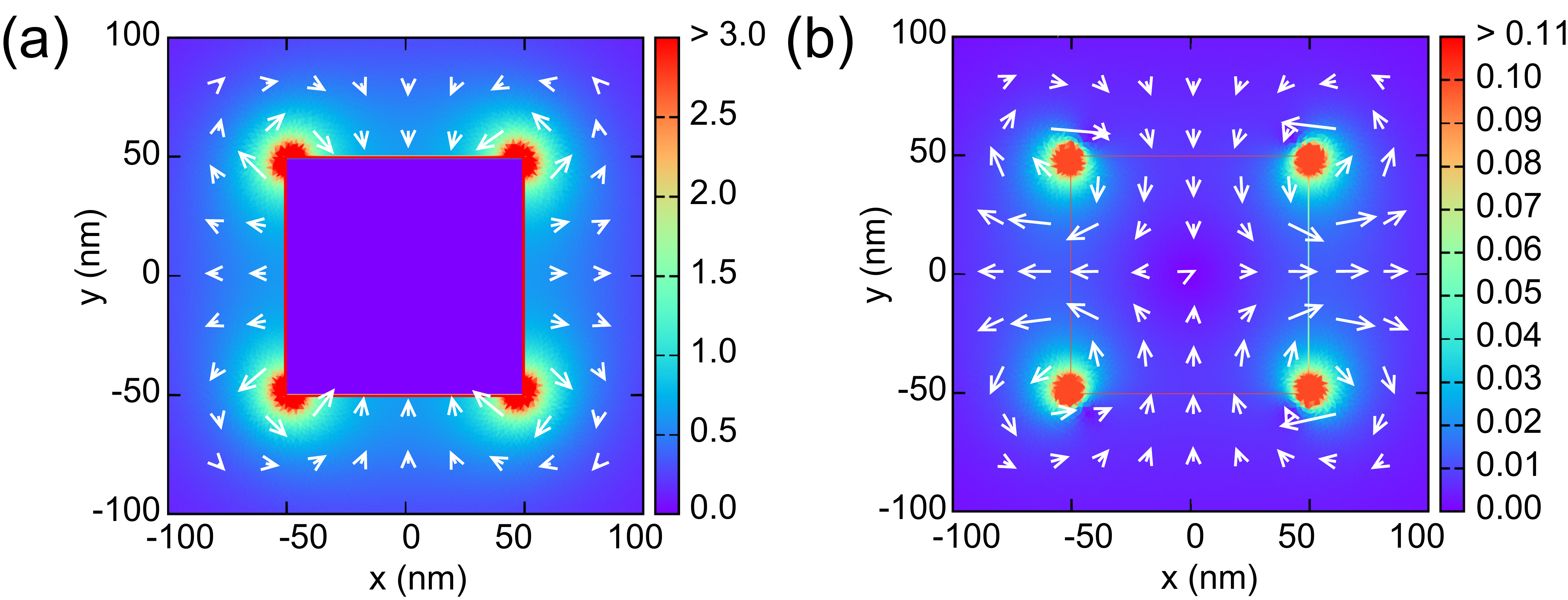}
\caption{\label{smoothing}
(a) The magnitude and the direction of the electric field (10$^7$ V/m) and (b) the magnetic induction (gauss) around the TI.}
\end{figure}

We lastly checked smoothing boundary conditions at corners in the model system (Fig. 4 in the main text).
The boundary conditions on the outer box are determined by the facet dependent work functions,
and there is a discrete jump at corners between two values.
In real nature the potential may change continuously so that we import the variation of the potential near the corners.
We set the potential value of the corner as the intermediate value of 
$(\phi_{\text{(111)}} + \phi_{\text{(}\bar{1}\text{10)}})/2$ = -5.44 V, and
the boundary values are gradually increasing (on the (111) surface) or decreasing (on the ($\bar{1}$10) surface)
on 5 nm region from the corner toward the surface center.
From this smoothing boundary conditions, we obviously obtained the same features as the previous results
with a little changing of the specific values.
The electric field and the magnetic induction are shown in Fig.~\ref{smoothing}, and at 5 nm away from the corner,
we obtain the electric field of 2.63$\times 10^7$ V/m and the magnetic induction
of 0.13 gauss. 

\bibliography{ti_prl}